\documentclass[journal=jacsat,manuscript=article]{achemso}

\usepackage[T1]{fontenc}
\usepackage{microtype}

\usepackage{amsmath}
\usepackage{amssymb}

\usepackage[version=4]{mhchem}

\usepackage{graphicx}
\graphicspath{{./}{figures/}}

\SectionNumbersOn
\setkeys{acs}{maxauthors=0}

\usepackage{xcolor}
\definecolor{revisionred}{RGB}{180,20,20}

\newif\ifshowrevisions
\showrevisionstrue

\ifshowrevisions
  
\else
  
\fi

\usepackage[
  breaklinks=true,
  colorlinks=true,
  allcolors=blue
]{hyperref}

\title{Nonadiabatic Dynamics near Multiple Light-Induced Conical Intersections under Bichromatic Driving: Quantum Wave-Packet Dynamics versus Floquet Surface Hopping}

\author{Jiayue Han}
\affiliation{Department of Chemistry, School of Science and Research Center for Industries of the Future, Westlake University, Hangzhou, Zhejiang 310030, China}

\author{Feng An}
\email{anfeng@westlake.edu.cn}
\affiliation{Department of Chemistry, School of Science and Research Center for Industries of the Future, Westlake University, Hangzhou, Zhejiang 310030, China}
\alsoaffiliation{Institute of Natural Sciences, Westlake Institute for Advanced Study, Hangzhou, Zhejiang 310024, China}

\author{Wenjie Dou}
\email{douwenjie@westlake.edu.cn}
\affiliation{Department of Chemistry, School of Science and Research Center for Industries of the Future, Westlake University, Hangzhou, Zhejiang 310030, China}
\alsoaffiliation{Institute of Natural Sciences, Westlake Institute for Advanced Study, Hangzhou, Zhejiang 310024, China}
\alsoaffiliation{Key Laboratory for Quantum Materials of Zhejiang Province, Department of Physics, School of Science and Research Center for Industries of the Future, Westlake University, Hangzhou, Zhejiang 310030, China}

\begin{document}

\begin{abstract}
Light-induced conical intersections (LICIs) create externally tunable pathways for nonadiabatic transitions, enabling active control of molecular photophysical and photochemical processes. However, both the dynamics near multiple LICIs under bichromatic driving and the applicability of our recently developed two-mode Floquet fewest switches surface hopping (two-mode F-FSSH) method to this regime remain insufficiently understood. Here, we construct a minimal three-channel Floquet Hamiltonian supporting two LICIs for \ce{Na2} interacting with a bichromatic field. We characterize its static Floquet properties and investigate the associated nonadiabatic dynamics using numerically exact quantum wave-packet dynamics and two-mode F-FSSH. We find that the second photon energy controls the relative positions of the LICIs, whereas the second-field intensity primarily redistributes and broadens the derivative-coupling landscape around the second LICI. Consequently, electronic population transfer and molecular alignment exhibit distinct and often nonmonotonic responses to these two control parameters. Two-mode F-FSSH reliably captures the principal features of the early-time dynamics and provides a semiquantitative description of the post-transient time-averaged observables. These findings advance our understanding of LICI-mediated dynamics at both the physical and methodological levels.
\end{abstract}

\noindent\textbf{TOC Graphic}

\begin{center}
\includegraphics[width=0.65\linewidth]{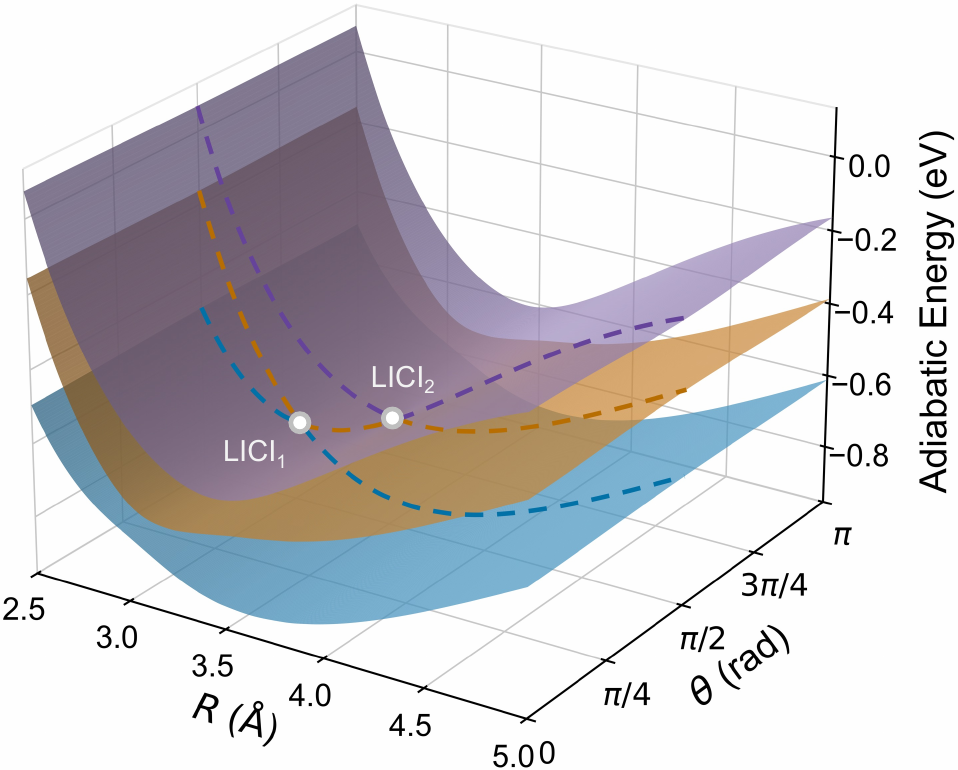}
\end{center}

\newpage

Conical intersections (CIs) occur when two adiabatic electronic states are energetically degenerate at a particular nuclear geometry. The corresponding potential energy surfaces locally exhibit a characteristic double-cone topology in the two-dimensional branching plane.\cite{Yarkony1996diabolical,Yarkony1998conical,Yarkony2001conical,Domcke2004conical,Domcke2011conical} When a nuclear wave packet passes through a CI region, it can rapidly split into components associated with different electronic states, resulting in population transfer between them.\cite{Worth2004beyond,Baer2006beyond,Levine2007isomerization,Takatsuka2011exploring,Zhu2016non,Schuurman2018dynamics} Vibronic coherence, quantum interference, and geometric-phase effects can further shape the nonadiabatic dynamics near CIs.\cite{Berry1984diabolical,Ryabinkin2017geometric,Valahu2023direct} These features make CIs central to photophysical and photochemical processes, including nonradiative excited-state relaxation and photoisomerization, thereby influencing molecular photostability, reaction selectivity, and quantum yields.\cite{Levine2006conical,Polli2010conical,Blancafort2014photochemistry,Farfan2020a}

Advances in extreme-ultraviolet time-resolved photoelectron spectroscopy (XUV-TRPES),\cite{von_Conta2018conical} time-resolved X-ray absorption spectroscopy (TRXAS),\cite{Neville2018ultrafast,Chang2020revealing} and two-dimensional electronic--vibrational (2DEV) spectroscopy\cite{Wu2019two} have enabled increasingly direct probes of nonadiabatic dynamics near CIs. More recently, growing attention has focused on the deliberate creation and manipulation of CIs with tailored laser fields, motivating the study of light-induced conical intersections (LICIs). In this context, Floquet theory provides a powerful framework for describing how periodic laser fields reshape molecular potential energy landscapes, with LICIs forming between Floquet quasienergy surfaces.\cite{Pawlak2015light,Fabri2024classical} Diatomic molecules are the simplest molecular systems capable of supporting LICIs. An isolated diatomic molecule possesses only one internal vibrational coordinate, the internuclear distance $R$, and therefore cannot support an ordinary field-free CI between electronic states of the same symmetry. Under linearly polarized driving, the angle $\theta$ between the molecular axis and the field-polarization direction enters the light--matter interaction as an additional dynamically relevant coordinate. Together, $R$ and $\theta$ span the two-dimensional branching plane required for LICI formation.\cite{Halasz2011light,Moiseyev2011the,Halasz2012light,Halasz2015direct,Natan2016observation,Szidarovszky2018direct,Fabri2022probing,Kiraly2026light} A seminal series of studies by Lorenz S.~Cederbaum and co-workers established \ce{Na2} as a prototypical system and laid the theoretical foundations for LICI-mediated dynamics in diatomic molecules.\cite{Moiseyev2008laser,Halasz2011conical,Moiseyev2011the,Sindelka2011strong,Halasz2011light,Halasz2012light,Halasz2012the,Szidarovszky2018direct} These studies provide a firm foundation for extending LICI research to more complex regimes of laser--molecule interaction.

Most of the aforementioned LICI studies focused on individual LICIs generated by monochromatic laser fields. Recent studies have begun to extend investigations of LICIs from monochromatic to bichromatic laser fields.\cite{Kubel2020probing,Sun2023directly,Sun2024manipulating,Barriga2024floquet} The additional photon-dressed channels introduced by bichromatic driving can generate multiple LICIs and offer new routes for controlling nonadiabatic dynamics. However, two key scientific questions still remain unresolved. From a physical perspective, how the photon energy and intensity of a secondary field jointly influence the coupled electronic and rovibrational dynamics near multiple LICIs has not been systematically established. From a methodological perspective, although fully quantum dynamical approaches have provided much of our current understanding, the ability of more tractable mixed quantum--classical approaches to reliably describe dynamics near multiple LICIs has not been benchmarked. Such mixed quantum--classical treatments are particularly valuable for extending LICI simulations to polyatomic molecules, which possess many more nuclear degrees of freedom and support coexisting intrinsic and light-induced CIs. As demonstrated in previous studies, surface hopping and its Floquet extensions represent promising options for this purpose.\cite{Ferretti1996quantum,Alijah1999fast,Malhado2014non,Humeniuk2016nonadiabatic,Fiedlschuster2017surface,Avanessian2024floquet,Avanessian2025intersystem} A natural next step is therefore a rigorous benchmark of our recently developed two-mode Floquet fewest switches surface hopping (two-mode F-FSSH) method.\cite{Han2026two} In this Letter, we revisit the \ce{Na2} molecule to investigate nonadiabatic dynamics near multiple LICIs generated by a bichromatic field. By combining numerically exact quantum wave-packet dynamics with two-mode F-FSSH, we aim to address the physical and methodological questions identified above.

We consider a \ce{Na2} molecule interacting with two continuous-wave laser fields linearly polarized along the same laboratory-fixed axis ($z$ axis). The two fields are characterized by photon energies $\hbar\omega_k$, peak electric-field amplitudes $E_k$, and cycle-averaged intensities $I_k$, with $k=1,2$. With both field phases set to zero, the total electric field is $\boldsymbol{\mathcal{E}}(t)=\mathbf{e}_z\sum_{k=1}^{2}E_k\cos(\omega_k t)$, where $\mathbf{e}_z$ is the unit vector along the common polarization direction. The field amplitude and intensity are related by $I_k=c\epsilon_0E_k^2/2$, where $c$ is the speed of light and $\epsilon_0$ is the vacuum permittivity. After separating the center-of-mass motion, the nuclear rotation is characterized by the rotational angular-momentum quantum number $J$ and its projection quantum number $M$ onto the space-fixed $z$ axis. Axial symmetry about the polarization axis makes $M$ a conserved quantity. Because the initially populated $J=0$ state necessarily has $M=0$, the dynamics remain within the $M=0$ subspace and are fully described by the internuclear distance $R$ and the molecular orientation angle $\theta$.

We retain only the $X\,{}^1\Sigma_g^+$ and $A\,{}^1\Sigma_u^+$ Born--Oppenheimer electronic states, hereafter referred to simply as the $X$ and $A$ states, respectively. The field-free potential energy curves $V_X(R)$ and $V_A(R)$ and the corresponding transition dipole moment $d(R)$ are taken from Refs.~\citenum{Zemke1981investigation} and~\citenum{Magnier1993potential}. We employ the truncated two-mode Floquet basis $\{|X,0,0\rangle,|A,-1,0\rangle,|A,0,-1\rangle\}$, where $|\alpha,n_1,n_2\rangle$ denotes a diabatic Floquet channel formed from electronic state $\alpha$ and the Floquet indices $n_1$ and $n_2$ associated with the two driving modes. A negative index denotes a Floquet replica shifted downward by one photon energy of the corresponding mode. In this basis, the effective three-channel Floquet Hamiltonian is
\begin{equation}
\begin{aligned}
\hat{H}^{\mathrm{F}}
&=
\hat{T}_{\mathrm{nuc}}\mathbf{I}_3
+
H_{\mathrm{el}}^{\mathrm{F}}(R,\theta)
\\[4pt]
&=
\left[
-\frac{\hbar^2}{2\mu}
\frac{\partial^2}{\partial R^2}
+
\frac{\hat{\mathbf{J}}^2}{2\mu R^2}
\right]\mathbf{I}_3
+
H_{\mathrm{el}}^{\mathrm{F}}(R,\theta),
\end{aligned}
\label{eq:floquet_total_hamiltonian}
\end{equation}
where $\mu$ is the reduced mass of \ce{Na2} and $\hat{\mathbf{J}}$ is the nuclear rotational angular-momentum operator. The nuclear kinetic-energy operator $\hat{T}_{\mathrm{nuc}}$ comprises the radial vibrational and end-over-end rotational kinetic-energy terms. The Floquet electronic Hamiltonian $H_{\mathrm{el}}^{\mathrm{F}}(R,\theta)$ has the explicit matrix form
\begin{equation}
H_{\mathrm{el}}^{\mathrm{F}}(R,\theta)
=
\begin{pmatrix}
V_X(R)
&
W_1(R,\theta)
&
W_2(R,\theta)
\\
W_1(R,\theta)
&
V_A(R)-\hbar\omega_1
&
0
\\
W_2(R,\theta)
&
0
&
V_A(R)-\hbar\omega_2
\end{pmatrix},
\label{eq:floquet_potential_matrix}
\end{equation}
with the field-induced couplings
\begin{equation}
W_k(R,\theta)
=
-\frac{E_k}{2}d(R)\cos\theta,
\qquad k=1,2.
\label{eq:floquet_coupling}
\end{equation}
Here $d(R)$ is the component of the body-fixed $X\leftrightarrow A$ transition dipole moment along the molecular axis. The two photon-dressed $A$-state channels are not directly coupled in the present model but interact indirectly through their common coupling to the $|X,0,0\rangle$ channel.

Unless otherwise specified, the calculations reported below use the following parameters. The photon energy and intensity of the first field are fixed at $\hbar\omega_1=1.968$~eV and $I_1=3.0\times10^{10}$~W\,cm$^{-2}$, respectively. Three target spatial arrangements of the LICIs are selected along $R$: the second LICI lies far to the right of the first, close to it on the right, or close to it on the left. The corresponding second-field photon energies required to realize these arrangements are $\hbar\omega_2=1.737$, $1.862$, and $2.095$~eV. For each value of $\hbar\omega_2$, the intensity ratio is varied over $I_2/I_1=0.2$, $1.0$, and $5.0$, giving the nine parameter combinations summarized in Table~\ref{tab:laser_parameters}.

\begin{table}[htbp]
\centering
\caption{Laser parameters used in the dynamical simulations. Each value of $\hbar\omega_2$ is combined with all three intensity ratios $I_2/I_1$. Here, $R_k$ is the internuclear distance at which the LICI associated with driving mode $k$ is located.}
\label{tab:laser_parameters}
\setlength{\tabcolsep}{5pt}
\renewcommand{\arraystretch}{1.15}
\begin{tabular}{cccccc}
\hline
$\hbar\omega_1$ (eV) &
$R_1$ (\AA) &
$\hbar\omega_2$ (eV) &
$R_2$ (\AA) &
$I_1$ (W\,cm$^{-2}$) &
$I_2/I_1$
\\
\hline
1.968 & 3.03 & 1.737 & 3.55 & $3.0\times10^{10}$ & 0.2, 1.0, 5.0
\\
1.968 & 3.03 & 1.862 & 3.25 & $3.0\times10^{10}$ & 0.2, 1.0, 5.0
\\
1.968 & 3.03 & 2.095 & 2.80 & $3.0\times10^{10}$ & 0.2, 1.0, 5.0
\\
\hline
\end{tabular}
\end{table}

\begin{figure}[t]
\centering
\includegraphics[width=0.6\linewidth]{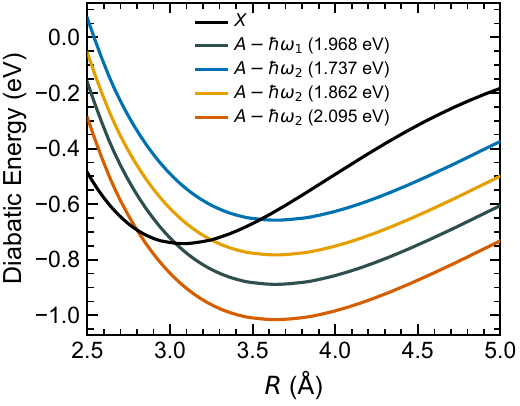}
\caption{Diabatic Floquet potential energy curves of \ce{Na2} in the effective three-channel model. The field-free potential $V_X(R)$ is shown together with the photon-dressed potentials $V_A(R)-\hbar\omega_1$ and $V_A(R)-\hbar\omega_2$, with $\hbar\omega_1=1.968$~eV and $\hbar\omega_2=1.737$, $1.862$, and $2.095$~eV.}
\label{fig:F-PESs_dia}
\end{figure}

Figure~\ref{fig:F-PESs_dia} shows the diabatic Floquet potential energy curves defined by the diagonal elements of Eq.~\eqref{eq:floquet_potential_matrix}. With $\hbar\omega_1$ fixed at $1.968$~eV, varying $\hbar\omega_2$ shifts the second photon-dressed $A$-state potential relative to $V_X(R)$ and thereby changes the crossing coordinate $R_2$.

The adiabatic Floquet quasienergies $\varepsilon_i(R,\theta)$ and states $|\phi_i(R,\theta)\rangle$ are obtained by solving
\begin{equation}
H_{\mathrm{el}}^{\mathrm{F}}(R,\theta)
|\phi_i(R,\theta)\rangle
=
\varepsilon_i(R,\theta)
|\phi_i(R,\theta)\rangle,
\qquad i=1,2,3.
\label{eq:floquet_diagonalization}
\end{equation}
Because the field-induced couplings $W_k(R,\theta)\propto\cos\theta$ vanish at $\theta=\pi/2$, each of the two diabatic crossings remains uncoupled at this angle. The effective three-channel model therefore contains two LICIs, located at $(R_1,\pi/2)$ and $(R_2,\pi/2)$. The associated quasienergy-gap and derivative-coupling landscapes are examined below.

\begin{figure}[t]
\centering
\includegraphics[width=\linewidth]{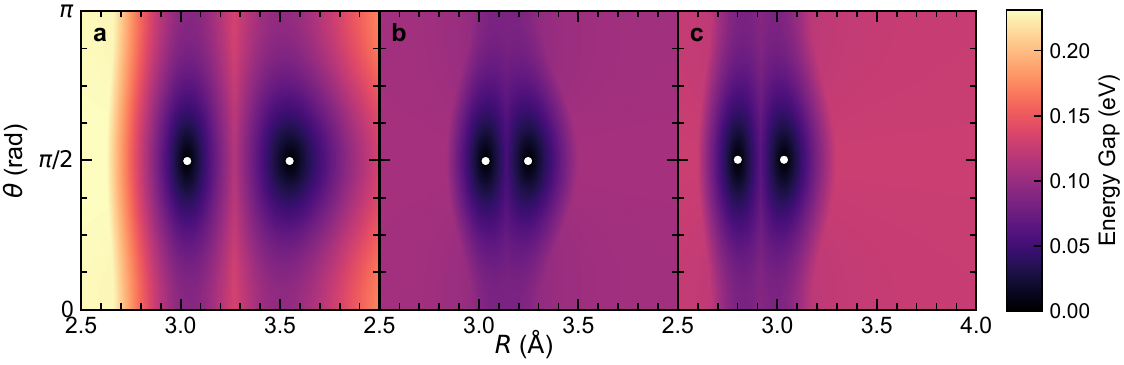}
\caption{Minimum gap between adjacent adiabatic Floquet quasienergy surfaces, $\Delta\varepsilon_{\min}=\min(\varepsilon_2-\varepsilon_1,\varepsilon_3-\varepsilon_2)$, as functions of the internuclear distance $R$ and molecular orientation angle $\theta$. The first-field intensity is fixed at $I_1=3.0\times10^{10}$~W\,cm$^{-2}$, with $I_2/I_1=1.0$. The photon energy of the first field is fixed at $\hbar\omega_1=1.968$~eV, whereas that of the second field is (a) $\hbar\omega_2=1.737$~eV, (b) $1.862$~eV, and (c) $2.095$~eV. The corresponding second LICI is located at (a) $R_2=3.55$~\AA, (b) $3.25$~\AA, and (c) $2.80$~\AA, whereas the first LICI remains at $R_1=3.03$~\AA. White circles mark the two LICI positions. Increasing the second photon energy shifts the second LICI toward smaller $R$ while leaving the first LICI fixed, thereby changing their separation and spatial ordering.}
\label{fig:energy_gap}
\end{figure}

Ordering the adiabatic Floquet quasienergies as $\varepsilon_1(R,\theta)\leq\varepsilon_2(R,\theta)\leq\varepsilon_3(R,\theta)$, we define the minimum gap between adjacent quasienergy surfaces as $\Delta\varepsilon_{\min}(R,\theta)=\min[\varepsilon_2-\varepsilon_1,\varepsilon_3-\varepsilon_2]$. Figure~\ref{fig:energy_gap} shows this quantity for the three values of $\hbar\omega_2$ at $I_1=3.0\times10^{10}$~W\,cm$^{-2}$ and $I_2/I_1=1.0$. Because $\hbar\omega_1$ is fixed, the first photon-dressed $A$-state potential and its crossing with $V_X(R)$ remain unchanged. Varying $\hbar\omega_2$ changes the energy offset of the second photon-dressed $A$-state potential relative to $V_X(R)$ and shifts their crossing position along $R$. The second photon energy therefore controls the position of the second LICI along $R$ and thereby the separation and spatial ordering of the two LICIs.

\begin{figure}[t]
\centering
\includegraphics[width=\linewidth]{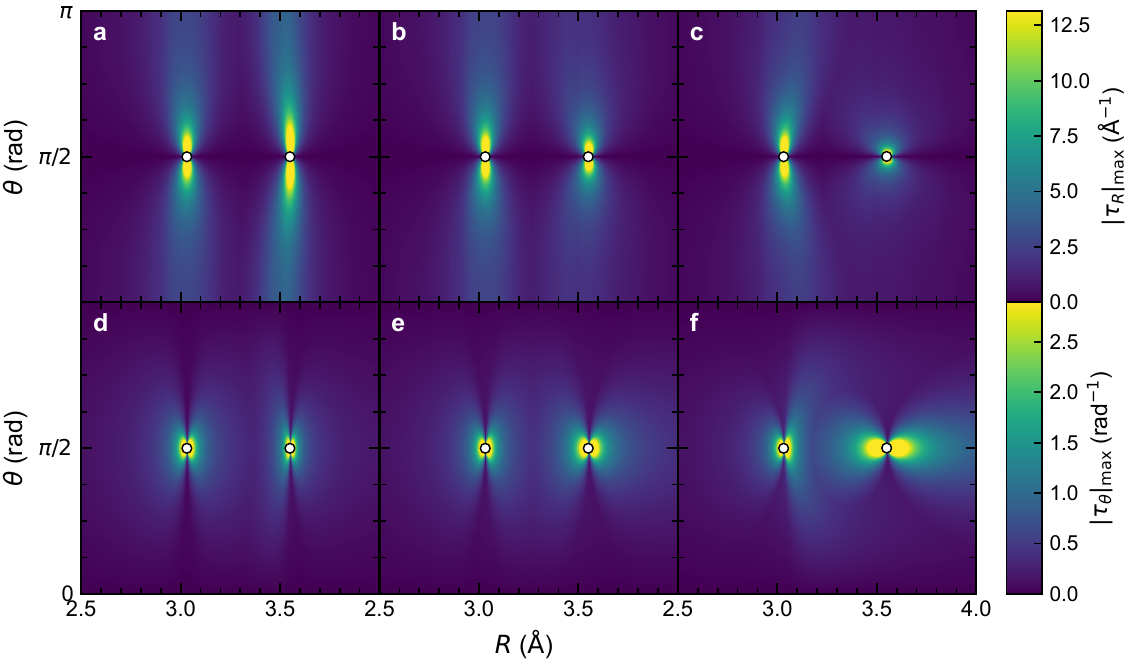}
\caption{Dominant derivative couplings between adjacent adiabatic Floquet states as functions of the internuclear distance $R$ and molecular orientation angle $\theta$. The photon energies are fixed at $\hbar\omega_1=1.968$~eV and $\hbar\omega_2=1.737$~eV, and the first-field intensity is $I_1=3.0\times10^{10}$~W\,cm$^{-2}$, placing the two LICIs at $R_1=3.03$~\AA\ and $R_2=3.55$~\AA, respectively. The columns correspond to (a,d) $I_2/I_1=0.2$, (b,e) $I_2/I_1=1.0$, and (c,f) $I_2/I_1=5.0$. The upper and lower rows show the dominant radial and angular components, respectively, selected from the derivative couplings between the adjacent-state pairs $(1,2)$ and $(2,3)$. White circles mark the two LICI positions. Increasing the second-field intensity redistributes and broadens the derivative coupling around the second LICI while leaving both LICI positions unchanged.}
\label{fig:derivative_coupling}
\end{figure}

The derivative coupling between adiabatic Floquet states $|\phi_i(R,\theta)\rangle$ and $|\phi_j(R,\theta)\rangle$ is resolved into radial and angular components, denoted by $\tau_{ij}^{R}(R,\theta)$ and $\tau_{ij}^{\theta}(R,\theta)$, respectively. Figure~\ref{fig:derivative_coupling} shows the dominant radial and angular derivative couplings between adjacent adiabatic Floquet states, defined at each nuclear configuration as
$|\tau_R|_{\max}=\max(|\tau_{12}^{R}|,|\tau_{23}^{R}|)$ and
$|\tau_\theta|_{\max}=\max(|\tau_{12}^{\theta}|,|\tau_{23}^{\theta}|)$, respectively. With the photon energies fixed at $\hbar\omega_1=1.968$~eV and $\hbar\omega_2=1.737$~eV and the first-field intensity fixed at $I_1=3.0\times10^{10}$~W\,cm$^{-2}$, changing $I_2/I_1$ leaves the two LICI positions unchanged but substantially reshapes the surrounding derivative-coupling landscape. As $I_2/I_1$ increases, both the radial and angular coupling features around the second LICI are redistributed and broadened, whereas those around the first LICI remain comparatively insensitive. The second-field intensity, by contrast, controls the spatial distribution and extent of the derivative coupling around the second LICI without changing either LICI position.

Together, these static Floquet properties show that the second photon energy controls the relative positions of the two LICIs, whereas the second-field intensity reshapes the derivative-coupling landscape around the second LICI. This separation of roles is consistent with earlier studies of LICIs generated by monochromatic fields.\cite{Moiseyev2008laser,Halasz2011conical,Halasz2011light,Moiseyev2011the,Sindelka2011strong,Halasz2012light,Halasz2012the,Halasz2015direct,Natan2016observation,Szidarovszky2018direct,Fabri2022probing,Kiraly2026light} Under bichromatic driving, however, the coexistence of two LICIs can give rise to more complex electronic and rovibrational dynamics. To characterize the nonadiabatic dynamics and evaluate the performance of two-mode F-FSSH, we compare numerically exact quantum wave-packet dynamics, hereafter referred to as ``Exact'', with two-mode F-FSSH. Further methodological and computational details are provided in the Computational Methods section at the end of the main text.

\begin{figure}[t]
\centering
\includegraphics[width=\linewidth]{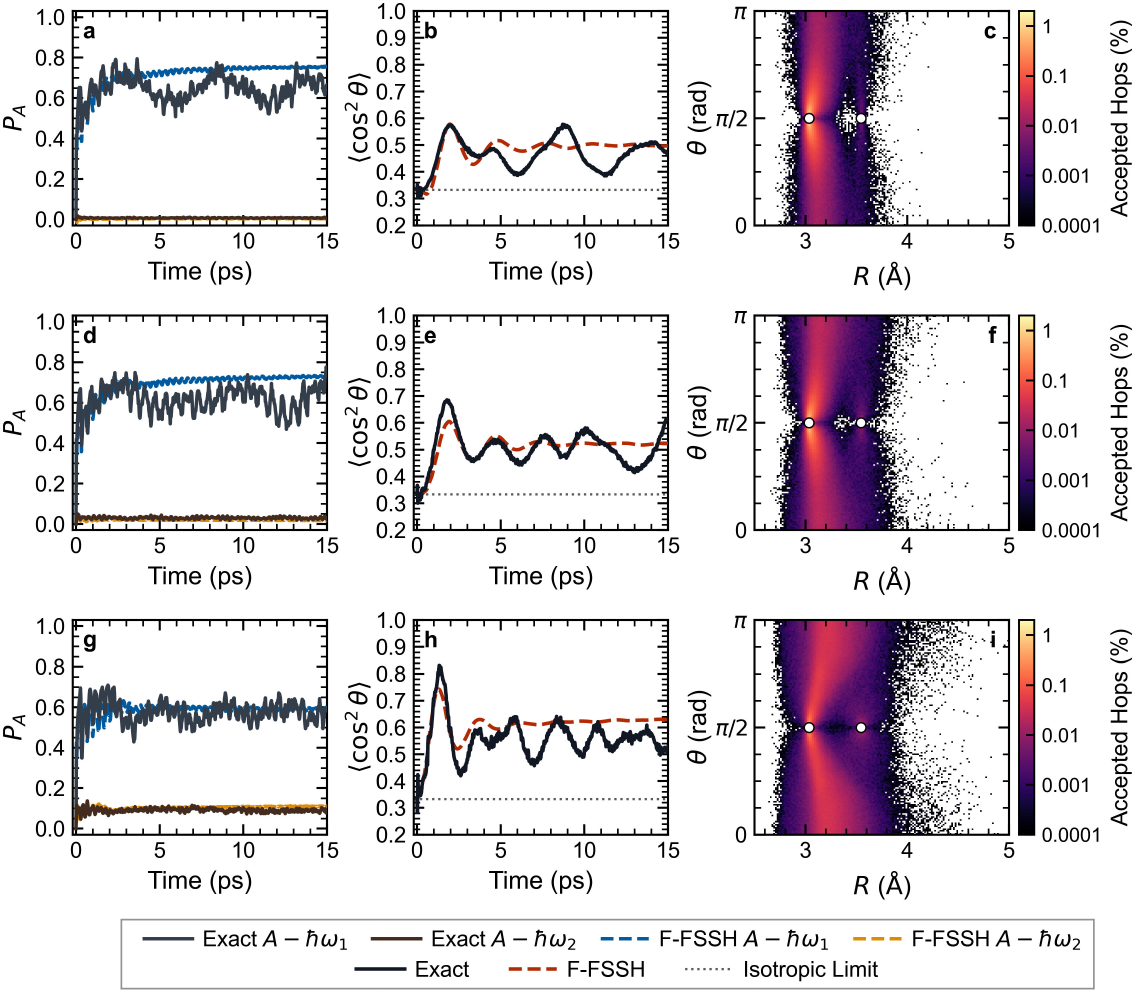}
\caption{Comparison of the nonadiabatic dynamics obtained using numerically exact quantum wave-packet propagation (Exact) and two-mode Floquet fewest switches surface hopping (two-mode F-FSSH) for $\hbar\omega_2=1.737$~eV. The first field is fixed at $\hbar\omega_1=1.968$~eV and $I_1=3.0\times10^{10}$~W\,cm$^{-2}$. The top (a--c), middle (d--f), and bottom (g--i) rows correspond to $I_2/I_1=0.2$, $1.0$, and $5.0$, respectively. The left column compares the diabatic Floquet-channel populations of $A_1\equiv|A,-1,0\rangle$ and $A_2\equiv|A,0,-1\rangle$, labeled $A-\hbar\omega_1$ and $A-\hbar\omega_2$, respectively. The middle column compares the molecular alignment $\langle\cos^2\theta\rangle$, with the horizontal dotted line marking the isotropic value $1/3$. The right column shows the spatial distributions of accepted hops obtained from two-mode F-FSSH in the $(R,\theta)$ plane. Each hop distribution is normalized independently by the total number of accepted hops and displayed on a logarithmic percentage scale. White circles mark the LICI positions.}
\label{fig:dynamics_1.737}
\end{figure}

\begin{figure}[t]
\centering
\includegraphics[width=\linewidth]{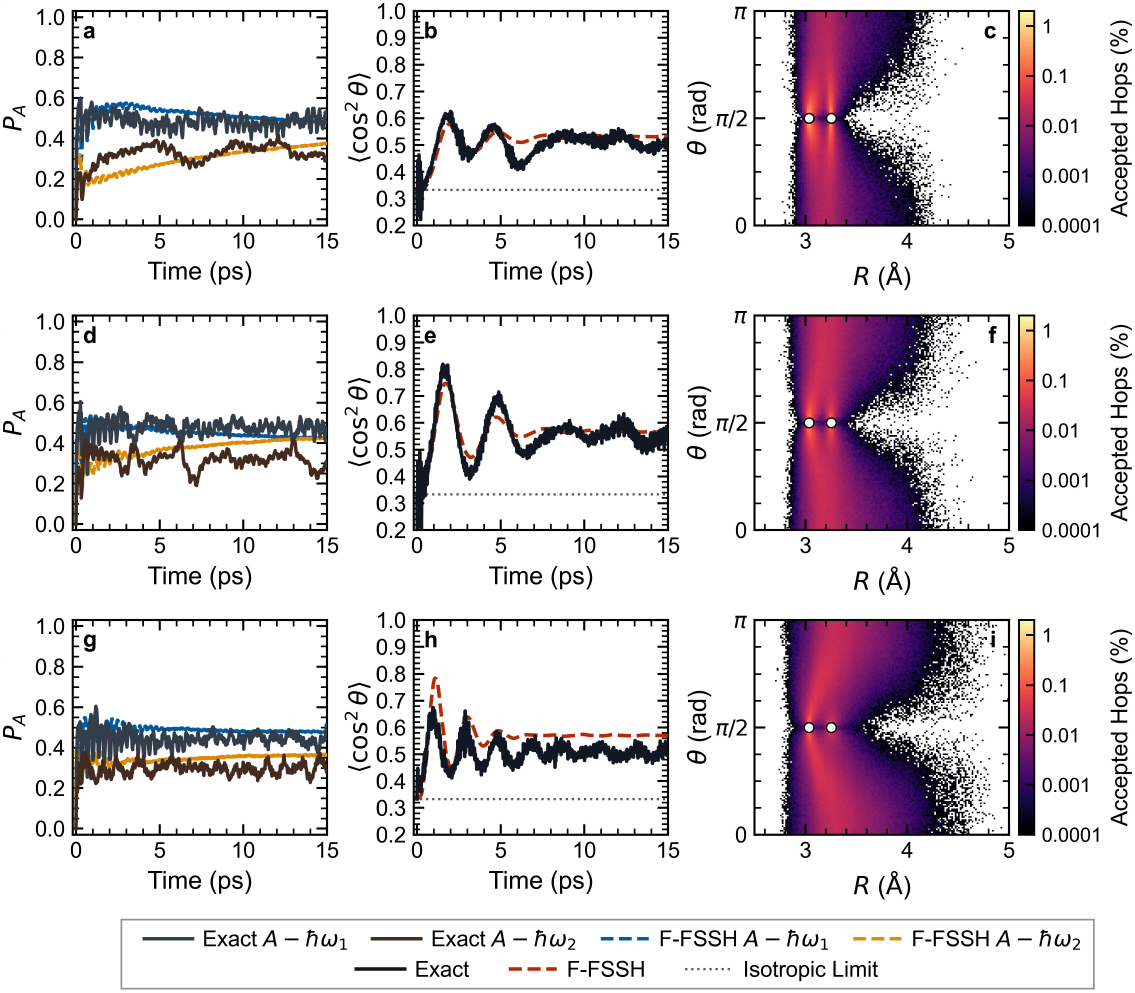}
\caption{Comparison of the nonadiabatic dynamics obtained using numerically exact quantum wave-packet propagation (Exact) and two-mode Floquet fewest switches surface hopping (two-mode F-FSSH) for $\hbar\omega_2=1.862$~eV. All remaining parameters, panel assignments, notation, and normalization are the same as in Figure~\ref{fig:dynamics_1.737}.}
\label{fig:dynamics_1.862}
\end{figure}

\begin{figure}[t]
\centering
\includegraphics[width=\linewidth]{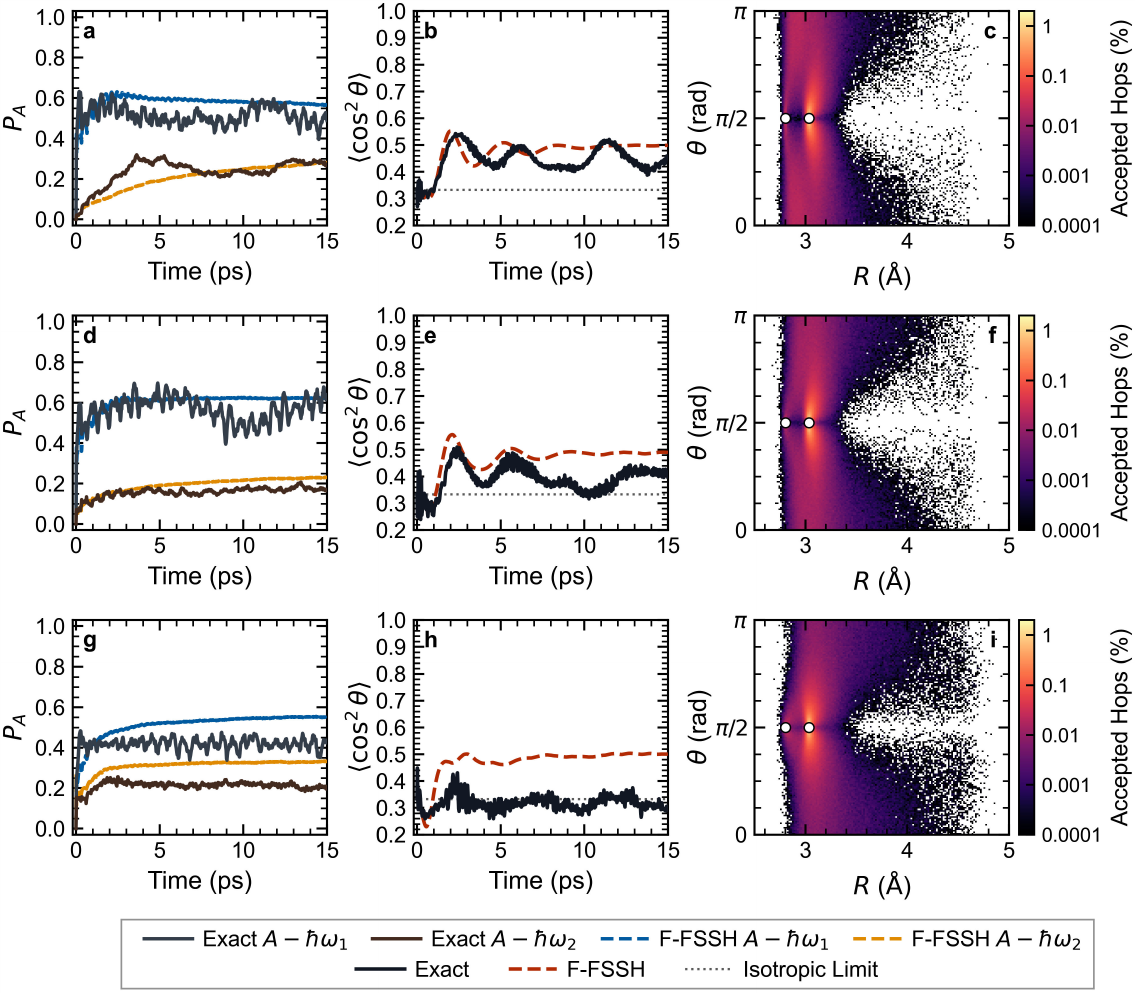}
\caption{Comparison of the nonadiabatic dynamics obtained using numerically exact quantum wave-packet propagation (Exact) and two-mode Floquet fewest switches surface hopping (two-mode F-FSSH) for $\hbar\omega_2=2.095$~eV. All remaining parameters, panel assignments, notation, and normalization are the same as in Figure~\ref{fig:dynamics_1.737}.}
\label{fig:dynamics_2.095}
\end{figure}

Figures~\ref{fig:dynamics_1.737}--\ref{fig:dynamics_2.095} compare the diabatic Floquet-channel populations $P_A$, molecular alignment $\langle\cos^2\theta\rangle$, and spatial distributions of accepted hops for the nine combinations of $\hbar\omega_2$ and $I_2/I_1$ listed in Table~\ref{tab:laser_parameters}.

For $\hbar\omega_2=1.737$~eV (Figure~\ref{fig:dynamics_1.737}), the two LICIs are well separated along $R$. The excited-channel population resides predominantly in $A_1\equiv|A,-1,0\rangle$ at all three intensity ratios. Increasing $I_2/I_1$ decreases the long-time $A_1$ population and increases the population of $A_2\equiv|A,0,-1\rangle$, although the latter remains the less populated channel. F-FSSH reproduces this intensity-dependent redistribution but overestimates the $A_1$ population at the two lower intensity ratios and suppresses the persistent oscillations present in the Exact populations. The alignment dynamics are considerably more sensitive to the second-field intensity. The Exact alignment remains above the isotropic value of $1/3$ after the initial transient, and its time-averaged magnitude increases monotonically with $I_2/I_1$. F-FSSH captures this enhancement but approaches a smoother long-time response and systematically overestimates the alignment, with the discrepancy becoming largest at $I_2/I_1=5.0$. Thus, at the lowest second photon energy, the second field can strongly modify the rotational dynamics even when its associated Floquet channel accounts for only a small fraction of the total excited-channel population.

For $\hbar\omega_2=1.862$~eV (Figure~\ref{fig:dynamics_1.862}), the second LICI lies closer to the first. The population is partitioned substantially more evenly between the two excited Floquet channels than at $\hbar\omega_2=1.737$~eV. Both excited Floquet channels remain appreciably populated, and the Exact populations exhibit persistent oscillations associated with repeated redistribution between them. Their relative partition depends only weakly on $I_2/I_1$. F-FSSH predicts a more pronounced intensity dependence and overestimates the $A_2$ contribution most clearly at $I_2/I_1=1.0$. Nevertheless, F-FSSH reproduces the coexistence of substantial populations in both channels and their overall long-time partition. The Exact alignment exhibits a nonmonotonic intensity dependence, reaching its largest time-averaged value at $I_2/I_1=1.0$ rather than under the strongest second field. Agreement with F-FSSH is close at the two lower intensity ratios, but F-FSSH overestimates the long-time alignment at $I_2/I_1=5.0$ and again replaces the persistent Exact oscillations with a smoother response.

For $\hbar\omega_2=2.095$~eV (Figure~\ref{fig:dynamics_2.095}), the second LICI lies to the left of the first. The excited-channel populations display a nonmonotonic dependence on $I_2/I_1$. In particular, the relative contribution of $A_2$ decreases from $I_2/I_1=0.2$ to $1.0$ and then increases again at $I_2/I_1=5.0$. F-FSSH captures this qualitative variation in the channel partition, but at the highest intensity it overestimates the absolute populations of both excited channels and yields substantially smoother dynamics. A more pronounced method dependence appears in the alignment response. The Exact time-averaged alignment decreases continuously as $I_2/I_1$ increases and falls slightly below the isotropic limit at $I_2/I_1=5.0$, indicating weak net anti-alignment. F-FSSH instead predicts persistent net alignment, $\langle\cos^2\theta\rangle>1/3$. At $I_2/I_1=5.0$, the discrepancy between the F-FSSH and Exact alignment dynamics is the largest among the nine parameter combinations. This result indicates that the strongly driven alignment dynamics in this regime are particularly sensitive to the approximations underlying the independent-trajectory treatment, although the present calculations do not identify a unique source of the discrepancy.

For all nine parameter combinations, we additionally use two-mode F-FSSH to map the spatial distributions of accepted hops in the $(R,\theta)$ plane. Each distribution is independently normalized by its total number of accepted hops and therefore represents the relative distribution of hopping events rather than their absolute number. The accepted hops are concentrated near $\theta=\pi/2$ and within the ranges of $R$ containing the two LICIs. Increasing $I_2/I_1$ generally increases the relative contribution of hops near the second LICI and broadens the region sampled around it.

\begin{figure}[t]
\centering
\includegraphics[width=0.9\linewidth]{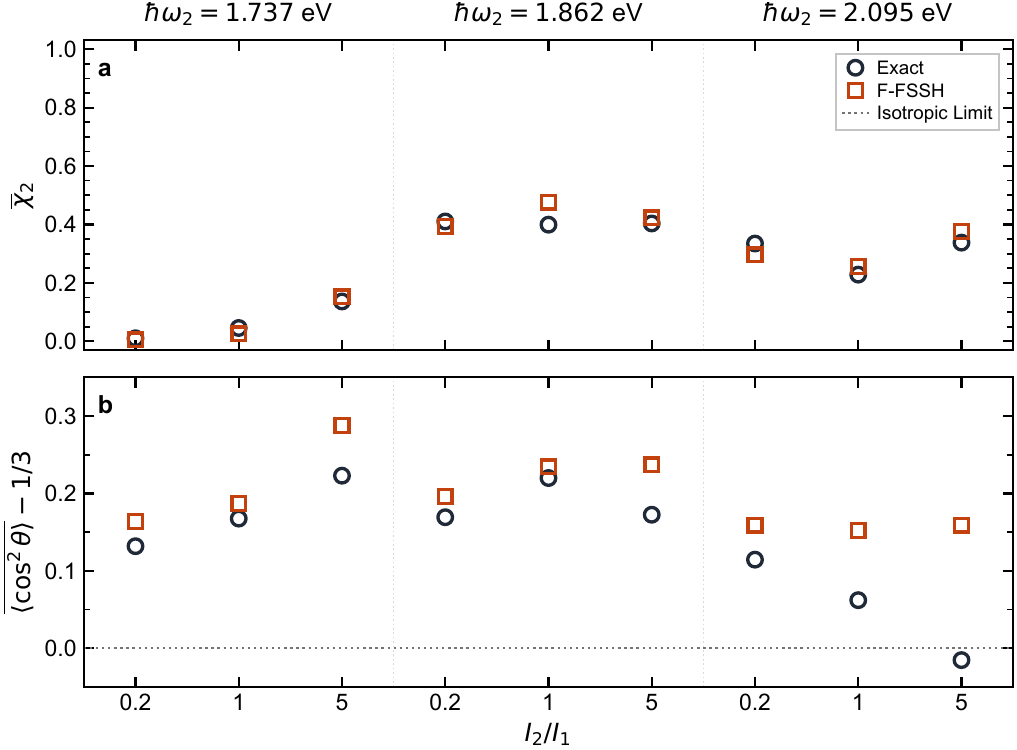}
\caption{Post-transient dynamical observables evaluated over $5$--$15$~ps for all nine field-parameter combinations. The three groups correspond to $\hbar\omega_2=1.737$, $1.862$, and $2.095$~eV, with $I_2/I_1=0.2$, $1.0$, and $5.0$ within each group. (a) Time-integrated second-channel fraction $\overline{\chi}_2$. (b) Time-averaged molecular alignment relative to the isotropic limit, $\overline{\langle\cos^2\theta\rangle}-1/3$. Circles and squares denote Exact and F-FSSH results, respectively. Vertical dotted lines separate the three groups of second-field photon energies, and the horizontal dotted line in panel (b) marks $\overline{\langle\cos^2\theta\rangle}-1/3=0$.}
\label{fig:dynamics_time_averaged}
\end{figure}

The persistent long-time oscillations make comparisons at individual times sensitive to the instantaneous oscillation phase. To obtain a compact characterization of the post-transient dynamics across the complete parameter set, we exclude the initial $5$~ps, which contain the principal population redistribution and alignment buildup, and analyze the subsequent interval $[t_1,t_2]=[5,15]$~ps. Using the channel labels defined above, we define the time-integrated fraction of the excited-channel population associated with $A_2$ as
\begin{equation}
\overline{\chi}_2
=
\frac{
\displaystyle\int_{t_1}^{t_2}P_{A_2}(t)\,dt
}{
\displaystyle\int_{t_1}^{t_2}
\left[P_{A_1}(t)+P_{A_2}(t)\right]dt
},
\label{eq:time_averaged_channel_fraction}
\end{equation}
where $P_{A_i}(t)$ denotes the population of the diabatic Floquet channel $A_i$. This conditional fraction characterizes the population partition between the two excited Floquet channels and is independent of their total population. The Time-averaged molecular alignment relative to the isotropic limit is defined as
\begin{equation}
\overline{\left\langle\cos^2\theta\right\rangle}
-\frac{1}{3}
=
\frac{1}{t_2-t_1}
\int_{t_1}^{t_2}
\left\langle\cos^2\theta\right\rangle(t)\,dt
-
\frac{1}{3}.
\label{eq:time_averaged_alignment}
\end{equation}
Positive and negative values of $\overline{\left\langle\cos^2\theta\right\rangle}
-1/3$ indicate net alignment and anti-alignment, respectively. Figure~\ref{fig:dynamics_time_averaged} summarizes these two post-transient observables for all nine field-parameter combinations.

Figure~\ref{fig:dynamics_time_averaged}a shows that the channel partition depends nonmonotonically on both the second photon energy and the second-field intensity. At $\hbar\omega_2=1.737$~eV, $\overline{\chi}_2$ remains small but increases monotonically with $I_2/I_1$. At $\hbar\omega_2=1.862$~eV, the Exact value remains close to $0.4$ over the full intensity range, indicating the most balanced and intensity-insensitive partition of the two excited channels. At $\hbar\omega_2=2.095$~eV, $\overline{\chi}_2$ decreases and then increases with $I_2/I_1$. For every intensity ratio, the largest second-channel fraction occurs at the intermediate photon energy. Thus, increasing either the second photon energy or the second-field intensity does not produce a universal monotonic change in the excited-channel population partition. F-FSSH reproduces these principal trends and the overall magnitude of $\overline{\chi}_2$.

Figure~\ref{fig:dynamics_time_averaged}b shows an even stronger dependence of the alignment response on the two control parameters. At $\hbar\omega_2=1.737$~eV, increasing $I_2/I_1$ progressively enhances the Exact alignment. At $1.862$~eV, the response is nonmonotonic and reaches its maximum at the intermediate intensity. At $2.095$~eV, increasing the intensity instead suppresses the alignment and ultimately produces weak anti-alignment. The effect of increasing the second-field intensity can therefore range from enhancement to a nonmonotonic response or quenching, depending on the position of the second LICI. F-FSSH consistently overestimates the time-averaged alignment relative to the Exact result. Its error increases at the higher intensity ratios and becomes particularly pronounced for $\hbar\omega_2=2.095$~eV, where it fails to reproduce the transition to anti-alignment.

Taken together, these time-dependent observables, their post-transient averages, and accepted-hop distributions characterize nonadiabatic dynamics near multiple LICIs under bichromatic driving. These results also provide a systematic benchmark of two-mode F-FSSH against the Exact calculations over a broad range of laser-field parameters. Several limitations of the present treatment, which also suggest directions for future methodological development, warrant further consideration.

The first limitation concerns nuclear coherence and repeated wave-packet passages. Because the relevant \ce{Na2} Floquet quasienergy surfaces are locally bound over the dynamically sampled range of $R$, the nuclear wave-packet components repeatedly return to the two LICI regions, where they can undergo successive splitting, recombination, and interference. The persistent oscillations in the Exact populations and alignment are consistent with such coherent recurrence. Standard F-FSSH faces two distinct difficulties in this regime. First, standard FSSH suffers from electronic overcoherence. Electronic amplitudes associated with different adiabatic states can remain coherent along an individual trajectory even after the corresponding nuclear wave-packet components would have spatially separated.\cite{Subotnik2016understanding,Miao2019revisiting,Jain2022pedagogical} Second, at the ensemble level, independent trajectories do not carry relative nuclear phases and therefore cannot reconstruct interference or recoherence when separated wave-packet components subsequently overlap. This limitation becomes increasingly apparent at longer times, when the F-FSSH populations and alignment evolve toward smooth profiles and fail to reproduce the persistent oscillations and recurrences present in the Exact results. The discrepancy can accumulate as the nuclear wave packet repeatedly passes through the LICI regions. Conventional decoherence corrections, such as the energy-based decoherence correction (EDC),\cite{Granucci2007critical,Granucci2010including} may partially mitigate intra-trajectory electronic overcoherence. However, such corrections cannot recover the relative nuclear phases between independent trajectories. Their overall effect on the accuracy of F-FSSH in this regime therefore remains unclear. Developing a general algorithm that reliably treats coherence in F-FSSH dynamics near single or multiple LICIs remains a substantial methodological challenge.

The second limitation concerns the nuclear Berry phase, which is not explicitly represented in standard F-FSSH. A nuclear wave packet acquires this topological phase upon encircling a LICI, potentially affecting the resulting nonadiabatic dynamics.\cite{Halasz2011light,Halasz2011conical,Ryabinkin2017geometric,Halasz2018geometric,Bouakline2018unambiguous,Valahu2023direct} Because the Exact calculation propagates the coupled nuclear wave packet in the diabatic Floquet representation, these Berry-phase effects are implicitly retained. Standard F-FSSH, by contrast, does not propagate a nuclear wave function and cannot explicitly represent the global phase difference between distinct nuclear paths. As discussed in the literature, Berry-phase effects may have only a limited influence during short-time passages through a CI, allowing surface-hopping methods that omit the Berry phase to remain reasonably accurate in this regime.\cite{Gherib2015why} This observation is consistent with the relatively good early-time agreement obtained here. Floquet phase-space surface hopping (F-PSSH) has been shown to incorporate Berry-phase effects for complex-valued Floquet Hamiltonians.\cite{Zhou2023nonadiabatic} Extending and testing this framework for the global geometric phase associated with multiple LICIs represents an important direction for future work.

A final limitation concerns the deliberately minimal truncation of the two-mode Floquet space. The present three-channel Floquet Hamiltonian should not be regarded as a Floquet-converged or physically complete representation of \ce{Na2} interacting with a bichromatic field. It nevertheless constitutes the minimal model containing two LICIs and isolates the associated nonadiabatic dynamics. Because the Exact and F-FSSH calculations employ the same $3\times3$ Hamiltonian, their comparison provides an internally consistent framework appropriate for the present methodological benchmark. A more physically complete Floquet description of the nonadiabatic dynamics of \ce{Na2} in a bichromatic field would require an enlarged Floquet space whose truncation is validated through systematic convergence tests. The resulting adiabatic Floquet quasienergy surfaces may contain more than two LICIs as well as numerous light-induced trivial crossings (LITCs), posing substantial numerical challenges for practical F-FSSH simulations.

In conclusion, we have investigated coupled electronic and rovibrational dynamics near two LICIs in \ce{Na2} under bichromatic driving using numerically exact quantum wave-packet dynamics and two-mode F-FSSH within a common effective three-channel Floquet Hamiltonian. The second photon energy controls the spatial separation and ordering of the two LICIs, whereas the second-field intensity primarily redistributes and broadens the derivative-coupling landscape around the second LICI. Their combined influence produces distinct and often nonmonotonic changes in electronic populations and molecular alignment, with no simple correspondence between these observables. The accepted-hop distributions obtained from two-mode F-FSSH show that increasing $I_2/I_1$ broadens the hopping distribution around the second LICI and increases the relative fraction of hops occurring there. Two-mode F-FSSH reliably reproduces the principal features of the early-time dynamics and provides a good description of the post-transient excited-channel partition $\overline{\chi}_2$. However, it smooths the persistent oscillations present in the Exact results and exhibits larger errors in the molecular alignment dynamics, particularly under strong second-field driving. Overall, this comparison delineates the applicability of two-mode F-FSSH to multiple-LICI dynamics and identifies nuclear coherence, geometric-phase effects, and Floquet-space convergence as priorities for further methodological development.

\section*{Computational Methods}

In this study, we employ two complementary methods: quantum wave-packet dynamics and mixed quantum--classical two-mode F-FSSH.

For the quantum wave-packet calculations, the radial and angular nuclear coordinates are represented using discrete variable representations (DVRs).\cite{Lill1982discrete,Light1985generalized,Colbert1992a,Light2000discrete} The nuclear wave packet $\boldsymbol{\Psi}(R,\theta,t)$ is propagated over each time step using the second-order symmetric split-operator approximation,\cite{Kosloff1983a,Kosloff1988time}
\begin{equation}
\boldsymbol{\Psi}(R,\theta,t+\Delta t)
=
\exp\left(
-\frac{i\Delta t}{2\hbar}\hat{T}_{\mathrm{nuc}}
\right)
\exp\left[
-\frac{i\Delta t}{\hbar}
H_{\mathrm{el}}^{\mathrm{F}}(R,\theta)
\right]
\exp\left(
-\frac{i\Delta t}{2\hbar}\hat{T}_{\mathrm{nuc}}
\right)
\boldsymbol{\Psi}(R,\theta,t)
+
\mathcal{O}(\Delta t^3).
\label{eq:split_operator}
\end{equation}
Within the effective three-channel Floquet Hamiltonian defined in Eq.~\eqref{eq:floquet_total_hamiltonian}, this grid-based propagation provides a numerically exact benchmark for two-mode F-FSSH.

The radial coordinate is represented by a 150-point sinc DVR over $R\in[0.5,6.5]$~\AA, and the angular coordinate is represented by a 35-point Gauss--Legendre DVR over $\theta\in[0,\pi]$. The propagation time step is $0.0242$~fs. The initial radial wave function is the numerically determined vibrational ground state of the field-free $X$-state potential, and the initial rotational wave function is the isotropic $J=0$ state. The wave packet is initialized entirely in the $|X,0,0\rangle$ channel. The diabatic Floquet-channel populations are evaluated directly from the corresponding wave-packet components, whereas the molecular alignment $\langle\cos^2\theta\rangle$ is evaluated from the total nuclear probability density.

For the two-mode F-FSSH calculations, we generalize the Cartesian-coordinate formulation introduced in Ref.~\citenum{Han2026two} to the curvilinear coordinates $(R,\theta)$. This extension enables two-mode F-FSSH to describe the coupled vibrational and rotational dynamics here. In the trajectory representation, the nuclear degrees of freedom are described by the canonical variables $(R,\theta,P_R,P_\theta)$. On the active adiabatic Floquet quasienergy surface $j$, these variables evolve according to Hamilton's equations,
\begin{equation}
\begin{aligned}
\dot{R}
&=
\frac{P_R}{\mu},
&
\dot{\theta}
&=
\frac{P_\theta}{\mu R^2},
\\[4pt]
\dot{P}_R
&=
-\frac{\partial\varepsilon_j(R,\theta)}{\partial R}
+
\frac{P_\theta^2}{\mu R^3},
&
\dot{P}_\theta
&=
-\frac{\partial\varepsilon_j(R,\theta)}{\partial\theta}.
\end{aligned}
\label{eq:fssh_nuclear_motion}
\end{equation}
Along each trajectory, the density matrix in the adiabatic Floquet basis evolves according to
\begin{equation}
\dot{\rho}_{mn}^{\mathrm{F}}
=
-\frac{i}{\hbar}
\left(\varepsilon_m-\varepsilon_n\right)\rho_{mn}^{\mathrm{F}}
-
\sum_l
\left[
\left(
\dot{R}\tau_{ml}^{R}
+
\dot{\theta}\tau_{ml}^{\theta}
\right)\rho_{ln}^{\mathrm{F}}
-
\rho_{ml}^{\mathrm{F}}
\left(
\dot{R}\tau_{ln}^{R}
+
\dot{\theta}\tau_{ln}^{\theta}
\right)
\right],
\label{eq:fssh_density_matrix}
\end{equation}
where $\rho_{mn}^{\mathrm{F}}$ denotes the corresponding density-matrix element between adiabatic Floquet states $m$ and $n$. The hopping probability from the active surface $j$ to target surface $k$ during a time step $\Delta t$ is
\begin{equation}
g_{j\rightarrow k}
=
\max
\left[
-\frac{2\Delta t}{\rho_{jj}^{\mathrm{F}}}
\operatorname{Re}
\left\{
\rho_{jk}^{\mathrm{F}}
\left(
\dot{R}\tau_{kj}^{R}
+
\dot{\theta}\tau_{kj}^{\theta}
\right)
\right\},
0
\right].
\label{eq:fssh_hopping_probability}
\end{equation}
When a hop is accepted, the conjugate momenta are adjusted along the normalized derivative-coupling direction according to
\begin{equation}
P_R^{\prime}
=
P_R
+
\kappa
\frac{\tau_{jk}^{R}}
{\sqrt{
\left|\tau_{jk}^{R}\right|^2
+
\left|\tau_{jk}^{\theta}\right|^2
}},
\qquad
P_\theta^{\prime}
=
P_\theta
+
\kappa
\frac{\tau_{jk}^{\theta}}
{\sqrt{
\left|\tau_{jk}^{R}\right|^2
+
\left|\tau_{jk}^{\theta}\right|^2
}}.
\label{eq:fssh_momentum_rescaling}
\end{equation}
The parameter $\kappa$ is determined by imposing conservation of the total energy,
\begin{equation}
\frac{\left(P_R^{\prime}\right)^2}{2\mu}
+
\frac{\left(P_\theta^{\prime}\right)^2}{2\mu R^2}
+
\varepsilon_k(R,\theta)
=
\frac{P_R^2}{2\mu}
+
\frac{P_\theta^2}{2\mu R^2}
+
\varepsilon_j(R,\theta).
\label{eq:fssh_energy_conservation}
\end{equation}
Among the real solutions of Eq.~\eqref{eq:fssh_energy_conservation}, the root with the smaller absolute value is selected to minimize the change in the nuclear momenta.

For each combination of field parameters, the two-mode F-FSSH results are averaged over $50,000$ independent trajectories. The propagation time step is $0.0484$~fs. The initial values of $R$ and $P_R$ are sampled from the Wigner distribution of the harmonic ground state obtained by expanding the field-free $X$-state potential about its equilibrium geometry. The initial molecular orientations are sampled isotropically by drawing $\cos\theta$ uniformly from $[-1,1]$, and $P_\theta$ is initialized to zero. This sampling represents an initially nonrotating isotropic ensemble, for which $u=\cos\theta$ is uniformly distributed over $[-1,1]$, yielding $\langle\cos^2\theta\rangle=1/2\int_{-1}^{1}u^2\,du=1/3$. The trajectories are initialized in the diabatic Floquet channel $|X,0,0\rangle$. The diabatic Floquet-channel populations are computed using the third method of Ref.~\citenum{Landry2013communication}. The molecular alignment is evaluated as the trajectory-ensemble average of $\cos^2\theta$. For each accepted surface hop, the nuclear coordinates are recorded to construct their spatial distributions.

\section*{Acknowledgments}

W.D.\ acknowledges financial support from the National Natural Science Foundation of China (Grant Nos.~22273075 and 22361142829) and the Zhejiang Provincial Natural Science Foundation (Grant No.~XHD24B0301). The authors acknowledge computational resources and technical support from the High-Performance Computing Center at Westlake University. J.H.\ thanks Dr.~Vahid Mosallanejad and Dr.~Yu Wang for helpful discussions.

\bibliography{references}

\end{document}